\documentclass[12pt]{article}

\usepackage{amsmath,amsfonts}
\usepackage{natbib}
\usepackage{float}
\usepackage{graphicx}
\usepackage{geometry}

\usepackage{amssymb}
\usepackage{bm}
\usepackage{psfrag,epsf}
\usepackage{enumerate}
\usepackage{multirow}
\usepackage{caption}
\usepackage{subcaption}
\usepackage{mathtools}
\usepackage{mathrsfs}
\usepackage{authblk}

\begin{document}

\title{\bf A Note on Using Discretized Simulated Data to Estimate Implicit Likelihoods in Bayesian Analyses}
 
\author[1]{M. S. Hamada}
\author[2]{T. L. Graves}
\author[1]{N. W. Hengartner}
\author[3]{D. M. Higdon}
\author[1]{A. V. Huzurbazar}
\author[1]{E. C. Lawrence}
\author[4]{C. D. Linkletter}
\author[5]{C. S.  Reese}
\author[6]{D. W. Scott}
\author[7]{R. R. Sitter}
\author[5]{R. L. Warr}
\author[1]{B. J. Williams}

\affil[1]{Los Alamos National Laboratory}
\affil[2]{Berry Consultants}
\affil[3]{Virginia Polytechnic Institute}
\affil[4]{The Lego Group}
\affil[5]{Brigham Young University}
\affil[6]{Rice University}
\affil[7]{Simon Fraser University}

 \maketitle

\begin{abstract}
This article presents a Bayesian inferential method where the likelihood for a model is unknown but where data can easily be simulated from the model. We discretize simulated (continuous) data to estimate the implicit likelihood in a Bayesian analysis employing a Markov chain Monte Carlo algorithm. Three examples are presented as well as a small study on some of the method's properties. 
\end{abstract}

\noindent{\sc KEY WORDS:  Gaussian process, interval-censored, Latin Hypercube Sample, Markov chain Monte Carlo (MCMC), probability, relative frequency}

\section{Introduction}\label{sec:intro}
This article considers the situation where the likelihood of the data is unknown, i.e., an implicit likelihood. However, we assume that it is easy to simulate data from the data model and that they can be used to estimate the likelihood, i.e., an estimated likelihood. We want to use the implicit likelihood estimate in a Bayesian analysis that incorporates prior knowledge. 

Approximate Bayesian Computation  or ABC 
\cite{Sisson2019} is the preeminent so-called ``likelihood-free" method today for handling implicit or intractable likelihoods and ABC also employs simulation. ABC compares simulated data summaries with observed data summaries to decide whether to accept the model parameters values used to generate the simulated data as samples from the posterior distribution. ABC relaxes the acceptance requirements that makes the sampling more efficient but results in samples drawn from an approximate posterior distribution.

\citet{Lerman1981} estimate probabilities of probability mass functions and
\citet{Diggle1984} use kernel density estimates (KDEs) in implicit likelihoods by simulating data
for maximum likelihood estimation.
\citet{Drovandi2015} presents Indirect Inference (II) methods that employ auxiliary models that are different from the data models, possibly even with different parameters, and makes connections to ABC.
Within II, 
\citet{Drovandi2015} views the use of KDEs as an Indirect Likelihood (IL) method of an auxiliary likelihood, i.e., the estimated likelihood where the summary statistics are the data.
\citet{Sisson2007} and 
\citet{Flury2011} also use sequential Monte Carlo (SMC) to handle implicit likelihoods within particle filters that rely on simulation.
\citet{Price2018} uses a multivariate normal approximation of an implicit likelihood in a method that is referred to as synthetic likelihood.

Despite 
\citet{Drovandi2015} making the connection between KDEs and ABC mentioned above, the
literature seems to rather silent on their use in Bayesian analyses. Perhaps the silence is because the dimension of the response needs to be small, say, less than 10, but the early applications motivating ABC had large dimensional responses.
Still in many engineering and physical sciences applications that we are involved with, the dimension of the response is small.
However, when we used a KDE embedded in a Markov chain Monte Carlo (MCMC) algorithm, we experienced several problems. With different bandwidth choices, either the MCMC samples did not mix well or they were seriously biased. Here, we applied KDEs in a brute force manner; that is, for every candidate parameter value, we simulated data, computed the KDE and used the KDE to evaluate the likelihood, i.e., the estimated likelihood. In this article, we consider discretizing the simulated (continuous) data and estimating the interval-censored probabilities by the simulated proportions. We then use these estimated probabilities to estimate the discretized data likelihood. We refer to the proposed method as SimILE for Simulated Implicit Likelihood Estimation for shorthand whereas the long name is discretized simulated data for implicit discretized data likelihood estimation.

The article is organized as follows. Section~\ref{sec:simlike} describes the SimILE method.
Section~\ref{sec:examples} presents three examples. Section~\ref{sec:study} explores some aspects of the SimILE method as well as using Gaussian process emulation as an alternative to the brute force use of simulation mentioned above. Section~\ref{sec:disc} concludes this article with a discussion.

\section{Discretizing Continuous Data and Estimated Likelihood}
\label{sec:simlike}

In this article, we propose discretizing continuous data so that the likelihood consists of probabilities of the discretized or interval-censored data. At some level, all continuous data are discrete because they are measured with error. Rounding is another example of discretization that can have little impact on inference, even for moderate amounts of rounding \cite{Hamada1989, Hamada1991}. 

An algorithm for the SimILE method is as follows:
\begin{description}
\item[I.] Calculate data-based intervals.
\begin{description}
\item[Ia.] Let $nInt$ be the target number of intervals.
\item[Ib.] Divide the range of the observed data ${\bf y}$ into $nInt+1$ evenly spaced intervals of width $w$, where $w=[{\rm max}({\bf y})- {\rm min}({\bf y})]/nInt$ and the first interval starts at ${\rm min}({\bf y})-w/2$ and the last interval ends at ${\rm max}({\bf y})+w/2$.
\item[Ic.] Add intervals before the first and last intervals if needed to cover the support of the distribution generating the data.
\end{description}
\item[II.] Calculate the discretized implicit likelihood.
\begin{description}
\item[IIa.] For a given set of parameter values, simulate $nSim$ data, $\tilde{y}_1, \ldots, \tilde{y}_{nSim}$.
\item[IIb.] Use the simulated data to compute the relative frequencies of the data-based intervals.
\item[IIc.] As in \cite{Flury2011}, 
the discretized implicit likelihood estimate is the product of the relative frequencies for the data-based intervals into which the observed data ${\bf y}$ fall.
\end{description}
\item[III.] For multivariate data, apply the same discretization simultaneously to all the coordinates.
\end{description}

\begin{figure}[H]
   \centerline{\includegraphics[width = 4.5in, height = 4.5in]{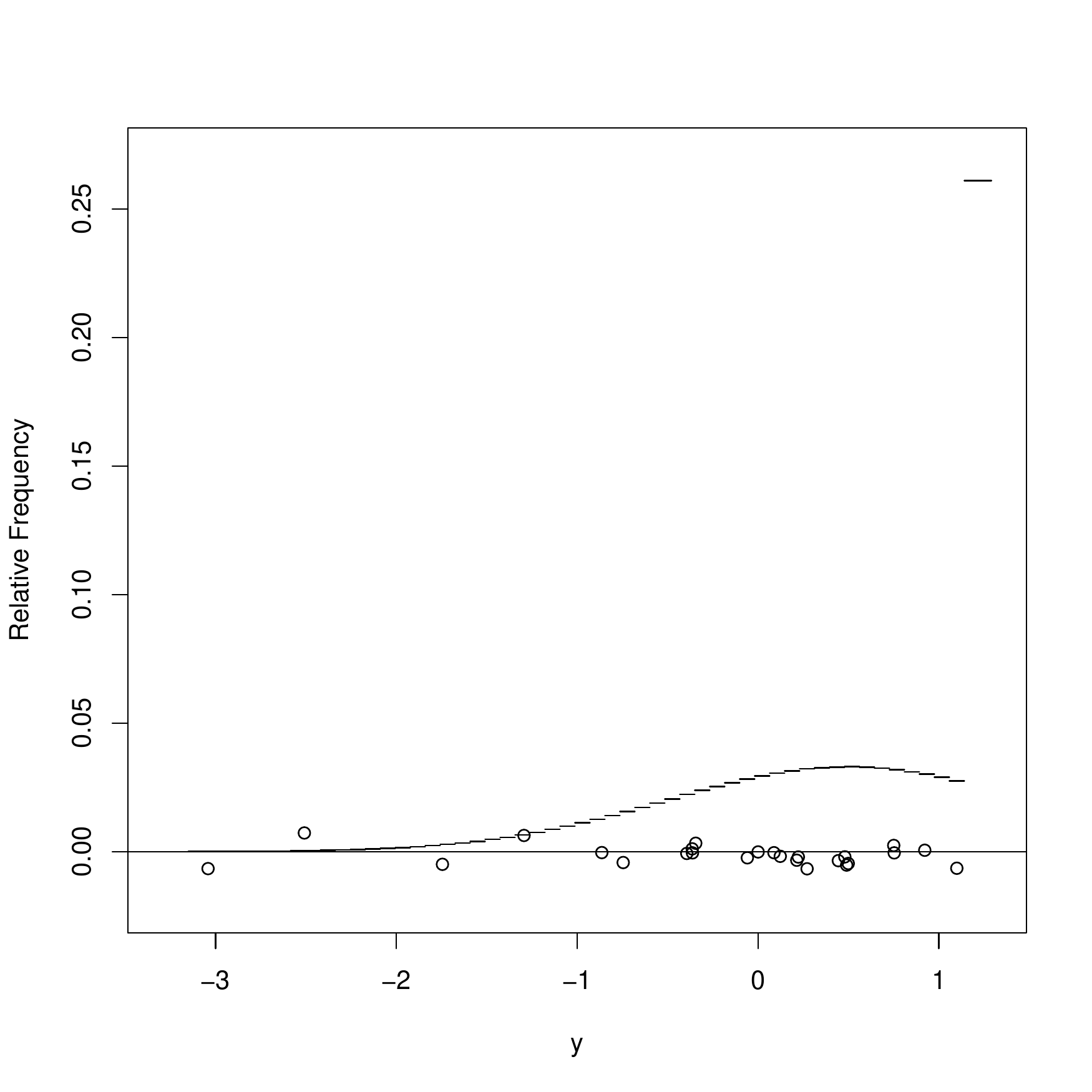}}
    \caption{Plot of relative frequencies and observed data (jiggled) for the normal example.}
    \label{fig:relFreq}
\end{figure}

\begin{figure}[H]
    \centering
       \includegraphics[width = 4.5in, height = 4.5in]{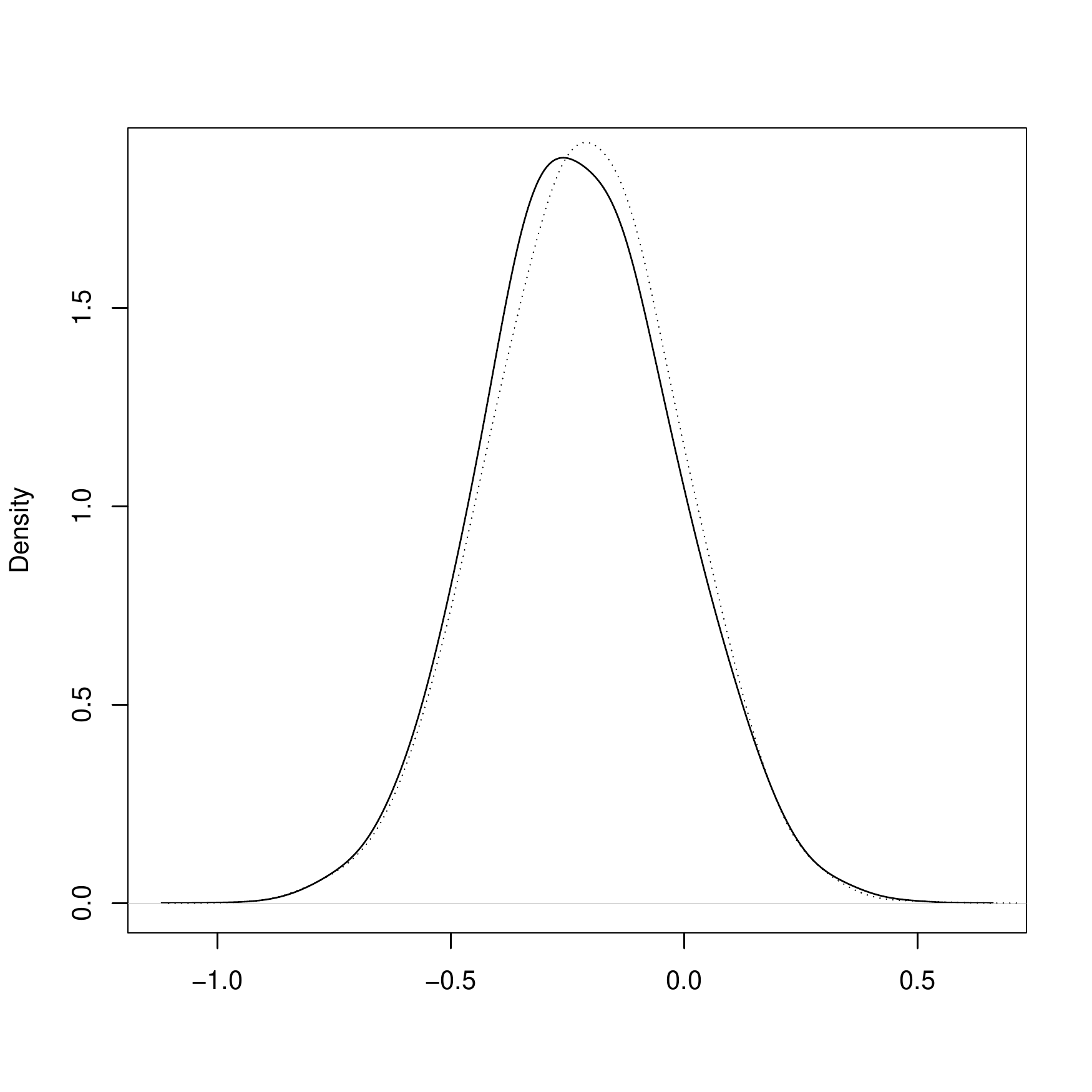}
    \caption{Plot of posterior distributions for the normal example, exact likelihood as the solid line and SimILE method as the dotted line.}
    \label{fig:posteriorsCompare}
\end{figure}

To help understand this algorithm better, 
consider a simple one parameter example of $n=25$ observed data generated by a $Normal(0, 1^2)$ distribution, where $\mu$ is the parameter of interest whose true value is $\mu=0$. We
use the prior $Normal(1,10^2)$ for $\mu$, and $nInt=50$ and $nSim=10^7$ in the algorithm. In Step Ib, the range of ${\bf y}$, the observed data of size 25, is $( -3.041,  1.099)$ so that $w=0.083 $; the start of the first interval is $-3.000$ and the end of the last interval is $1.140$. Since the support of the normal distribution is the real line, we add intervals $(-\infty, -3.000)$ and $(1.140, \infty)$ in Step Ic. Suppose that we evaluate the discretized implicit likelihood estimate at $\mu=0.5$. Then, from one realization of $\tilde{y}_1, \ldots, \tilde{y}_{nSim}$ in Step IIa, the relative frequencies in Step IIc are
0.0002348, 0.0000846, $\ldots$, 0.0276046, 0.2611034; see the plot of the relative frequencies and the observed data (jiggled) in Figure~\ref{fig:relFreq}. Based on Step IIc, the discretized implicit likelihood estimate
loglikelihood is the sum of the log of the relative frequencies of the data-based intervals into which the observed data ${\bf y}$ fall is $-106.274$.
We use the MCMC algorithm (i.e., Metropolis-Hastings) with the exact and discretized implicit likelihood estimate to draw 11,000 samples from the posterior distribution of $\mu$.
After dropping the first 1,000 draws as burnin, Table~\ref{tab:normalExample} summarizes the subsequent 10,000 draws. We see  that the exact likelihood and discretized implicit likelihood estimate produce very similar results. We also see the similarity in the display in their posterior distributions in Figure~\ref{fig:posteriorsCompare}.

\begin{table}[H]
    \center
   \caption{Posterior Summaries for  $\mu$ for the  Normal Example}
    \label{tab:normalExample}
     \vspace{0.05 in}
    \renewcommand{\baselinestretch}{1}
\large\normalsize
    \begin{tabular}{lrrrrr}
        \hline
Method &Mean&Std. Dev. & 2.5\% & 50\% & 97.5\%\\
 \hline
Exact Likelihood& -0.224 &  0.201 & -0.612 & -0.229&  0.174\\
SimILE&   -0.213 & 0.199 & -0.607& -0.213 &  0.168 \\
        \hline  
    \end{tabular}
    \renewcommand{\baselinestretch}{1.5}
\large\normalsize
\end{table}

As mentioned in Section~\ref{sec:intro},
we tried to treat the data directly as continuous by simulating data and using kernel density estimates of probability density functions to obtain implicit likelihood estimates. However, for MCMC algorithms, our experience found that recommended bandwidths did not work, yielding either biased results or the MCMC algorithms getting stuck and not mixing properly. Moreover, when the bandwidths were shortened to reduce the bias, the MCMC algorithms also got stuck and and did not mix properly. 
On the other hand,
the SimILE method estimates  the discretized (interval-censored) probabilities by unbiased simulated relative frequencies. Note that
\citet{Flury2011} cites \cite{Andrieu2010} 
who shows how an unbiased estimate of the likelihood embedded in an MCMC algorithm provides proper draws from the posterior distribution.

\section{Examples}
\label{sec:examples}

In this section, we demonstrate the proposed SimILE method with three examples. The first arises when indirect measurements are collected. The second illustrates a situation where there is a mixture of implicit and explicit likelihoods. The third involves bivariate data.

\subsection{Indirect Measurements}

Consider a situation in which a population of maximum distances $x$ is of interest that follow a specified distribution. A measurement is taken at a randomly chosen angle $\theta$ of the part so that the measured distance $y$ is an attenuation of the maximum distance $x$ and is related by $y=x|{\rm{sin}(\theta})|$; we refer to $y$ as an indirect measurement. Here, we assume that the maximum distances $x$ follow a $Lognormal(0,1)$ distribution with $\mu=0$ and $\sigma=1$.

Such indirect measurements are easily simulated by draws of lognormally distributed maximum distances and pairing them with draws of uniformly distributed angles. Thereby, the proposed SimILE method can be applied.

\begin{figure}[H]
    \centering
   \includegraphics[width = 6in, height = 3.5in]{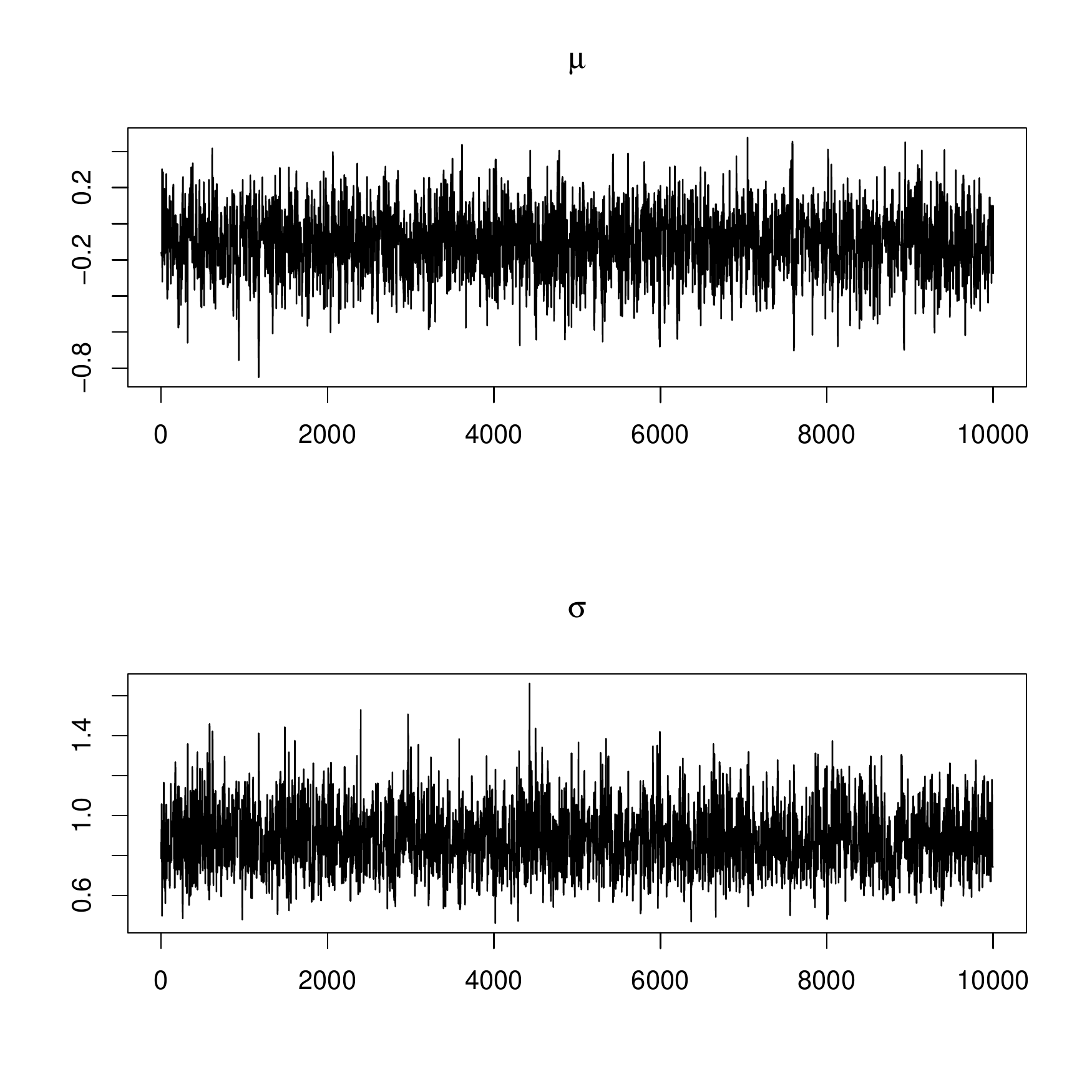}
    \caption{SimILE method posterior draws of $\mu$ and $\sigma$ for indirect measurements example. }
    \label{fig:implicitPost}
\end{figure}

We simulated a data set of size $N$ with $N=50$, where the true parameter values are $\mu=0$ and $\sigma=1$. The SimILE method was applied using
$nInt=100$ and $nSim=10^7$ with the following priors: $\mu \sim Normal(0,10)$ and $\sigma \sim Lognormal(0,10)$.
To obtain posterior samples, we use the Metropolis-Hastings algorithm \cite{Chib1995, Gelman2014}; 
we discard the first 1,000 samples as burnin and report on results from 10,000 subsequent samples. See Figure~\ref{fig:implicitPost} of the trace of the posterior draws of $(\mu, \sigma)$ that shows that they are mixing well.
The results for the SimILE method are given in Table~\ref{tab:implicitPosterior}.

\begin{table}[H]
    \center
   \caption{Posterior Summaries for the Implicit Measurement Example}
    \label{tab:implicitPosterior}
     \vspace{0.05 in}
    \renewcommand{\baselinestretch}{1}
\large\normalsize
    \begin{tabular}{llrrrrr}
        \hline
Method &Parameter&Mean&Std. Dev. & 2.5\% & 50\% & 97.5\%\\
        \hline
SimILE&$\mu$ &  -0.104 &  0.172 & -0.449  & -0.104 &  0.215\\
& $\sigma$ &  0.881 & 0.151 & 0.611 & 0.871 & 1.120 \\
        \hline
                \hline
Exact&$\mu$ &  -0.110  & 0.178 & -0.472 & -0.108  & 0.223\\
Likelihood& $\sigma$ &  0.892 & 0.165 & 0.602&  0.879 & 1.246 \\

                \hline
    \end{tabular}
    \renewcommand{\baselinestretch}{1.5}
\large\normalsize
\end{table}

It turns out that in this example, the pdf of $y$ can be expressed as an integral \cite{Graves2005}. 
See the results using the exact likelihood in Table~\ref{tab:implicitPosterior}. The results for the SimILE method are nearly the same.

\subsection{A Flowgraph with a Mixture of Implicit and Explicit Likelihoods}

\begin{figure}[H]
    \centering
      \includegraphics[height = 1.6in]{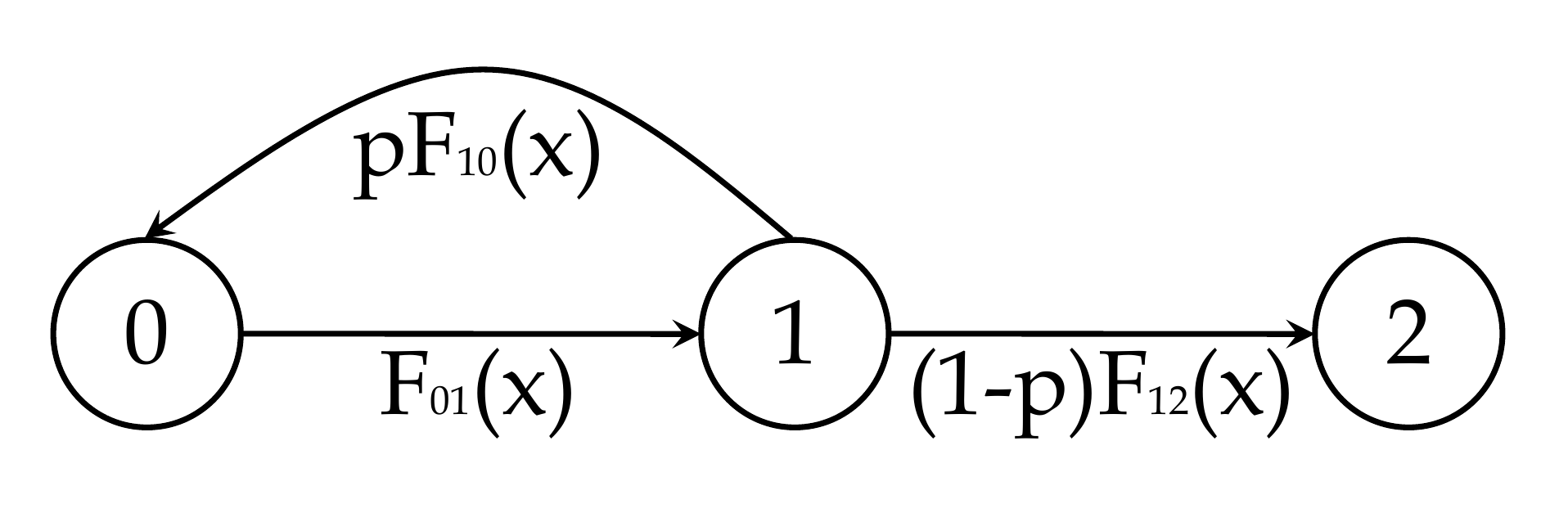}
    \caption{Flowgraph for transitions from state 0 to state 1, the return back to state 0 and to final state 2. }
    \label{fig:flowgraph}
\end{figure}

Figure~\ref{fig:flowgraph} displays a simple flowgraph of a process 
\cite{Huzurbazar2005} that can be in one of three states (0, 1, 2) such as working, degraded and failed. 
Starting in state 0, the transition time to state 1 is described by the cumulative distribution function (cdf) $F_{01}(x)$.
With probability $p$, there is a transition back to state 0 in time described by the cdf $F_{10}(x)$. Finally, a transition from state 1 to state 2, the final state, occurs with probability $1-p$ in time described by the cdf $F_{12}(x)$.\\

Consider the situation where we have transition times for the three transitions, 0 to 1, 1 to 0, and 1 to 2. The transition times provide information about the three transition time distributions. Here, we assume that the transition time distributions are gamma distributions parameterized in terms of their means and variances denoted generically by $\mu$ and $\sigma^2$, respectively. 
From the observed transitions, we can also estimate $p$ with the proportion of transitions from state 1 to state 0.
Independently, we have transition times from state 0 to state 2 in which the number of returns to state 0 is unknown; these data provide information about the model parameters from the three transition time distributions and $p$.

In this example, there are five components to the likelihood: the transition times for 0 to 1, 1 to 0, and 1 to 2, the binomial count $x$, the number of transitions from 1 to 0 out of $n$ transitions from state 1, and the 0 to 2 transition times. The first three are products of gamma probability density functions (pdfs) and the fourth is a binomial probability. The fifth is a product of implicit pdfs. That is, this example illustrates a situation where both implicit and explicit pdfs that contribute to the likelihood. 
We simulated data using 
$\mu_{01} = 6$,
$\sigma^2_{01} = 136/13$,
$\mu_{10}  = 4$,
$\sigma^2_{10} = 16/7$,
$\mu_{12} = 10$,
$\sigma^2_{12} = 100/12$, and
$p = 0.4$. There were 28 state 0 to 1 transition times, 11 state 1 to 0 transition times, and 17 state 1 to 2 transition times, with $x=11$ out of $n=28$. Independent of these data, we also have an 25  state 0 to state 2 transition times.

We can use the SimILE method for the state 0 to 2 transition time pdfs, in which use $nInt=100$ and $nSim=10^7$. For the other data, we use the exact likelihoods. Priors for $\mu_{01}$, $\sigma^2_{01}$, $\mu_{10}$, $\sigma^2_{10}$, $\mu_{12}$, $\sigma^2_{12}$,  $p$: 
$\mu_{01} \sim Lognormal(\log(6), 3)$, $\sigma^2_{01} \sim Lognormal(\log(136/13), 3)$,
$\mu_{10} \sim Lognormal(\log(4), 3)$, $\sigma^2_{10} \sim Lognormal(\log(16/7), 3)$,
$\mu_{12} \sim Lognormal(\log(10), 3)$, $\sigma^2_{12} \sim Lognormal(\log(100/12), 3)$,
and
${\rm logit}(p) \sim Normal(0,3)$.
To obtain posterior samples, we use the Metropolis-Hastings algorithm;
we discard the first 1,000 samples as burnin and report the results from 10,000 subsequent samples. Plots of posterior draws of the seven model parameters, not shown here, display good mixing. Table~\ref{tab:flowgraphPosterior} presents posterior summaries for the seven model parameters: $\mu_{01}$, $\sigma^2_{01}$, $\mu_{10}$, $\sigma^2_{10}$, $\mu_{12}$, $\sigma^2_{12}$,  $p$.

\begin{table}[H]
    \center
   \caption{Posterior Summaries for the Flowgraph Example}
    \label{tab:flowgraphPosterior}
     \vspace{0.05 in}
    \renewcommand{\baselinestretch}{1}
\large\normalsize
    \begin{tabular}{llrrrrr}
        \hline
Method &Parameter&Mean&Std. Dev. & 2.5\% & 50\% & 97.5\%\\
 \hline

SimILE & $\mu_{01}$ &  6.053 & 0.305 & 5.480 & 6.047 & 6.666\\  
 & $\sigma^2_{01}$ &   2.864 & 0.851  & 1.653 & 2.718  & 4.966 \\ 
    & $\mu_{10}$ &   3.658 & 0.328  & 3.069  & 3.638  & 4.343\\ 
    & $\sigma^2_{10}$ &  1.324 & 0.748 & 0.516 &   1.144 & 3.166  \\ 
    & $\mu_{12}$   & 8.602 & 0.436 & 7.770 & 8.580 & 9.501\\ 
    & $\sigma^2_{12}$ &   4.307  & 1.584 & 2.245  & 3.962 & 8.311 \\ 
    & $p$ &  0.412 & 0.059  & 0.296 & 0.412 & 0.525  \\ 
                    \hline
                            \hline
        Exact & $\mu_{01}$ & 6.043 & 0.291 & 5.495 & 6.036 & 6.640 \\ 
Likelihood & $\sigma^2_{01}$ & 2.840 & 0.876 & 1.610 & 2.682 & 4.855 \\ 
 &$\mu_{10}$ & 3.670 & 0.339 & 3.071 & 3.649 & 4.379 \\ 
 &$\sigma^2_{10}$ & 1.372 & 0.879 & 0.519 & 1.141 & 3.723 \\ 
 &$\mu_{12}$ & 8.603 & 0.437 & 7.762 & 8.588 & 9.467 \\ 
 &$\sigma^2_{12}$ & 4.253 & 1.517 & 2.224 & 3.955 & 8.046 \\ 
 &$p$ & 0.414 & 0.057 & 0.303 & 0.414 & 0.527 \\ 

        \hline
    \end{tabular}
    \renewcommand{\baselinestretch}{1.5}
\large\normalsize
\end{table}

\citet{Warr2010} show how the exact pdf of the 0 to 2 transition time can be numerically evaluated. See the results using the exact likelihood in Table~\ref{tab:networkPosterior}. The results for the SimILE method are nearly the same. This example shows how the practitioner who does not have code to evaluate the exact pdf of the 0 to 2 transition time can still perform an analysis using the proposed SimILE method.

\subsection{A Network With Bivariate Data}

\begin{figure}[H]
    \centering
   \includegraphics[width = 4.5in, height = 4.5in]{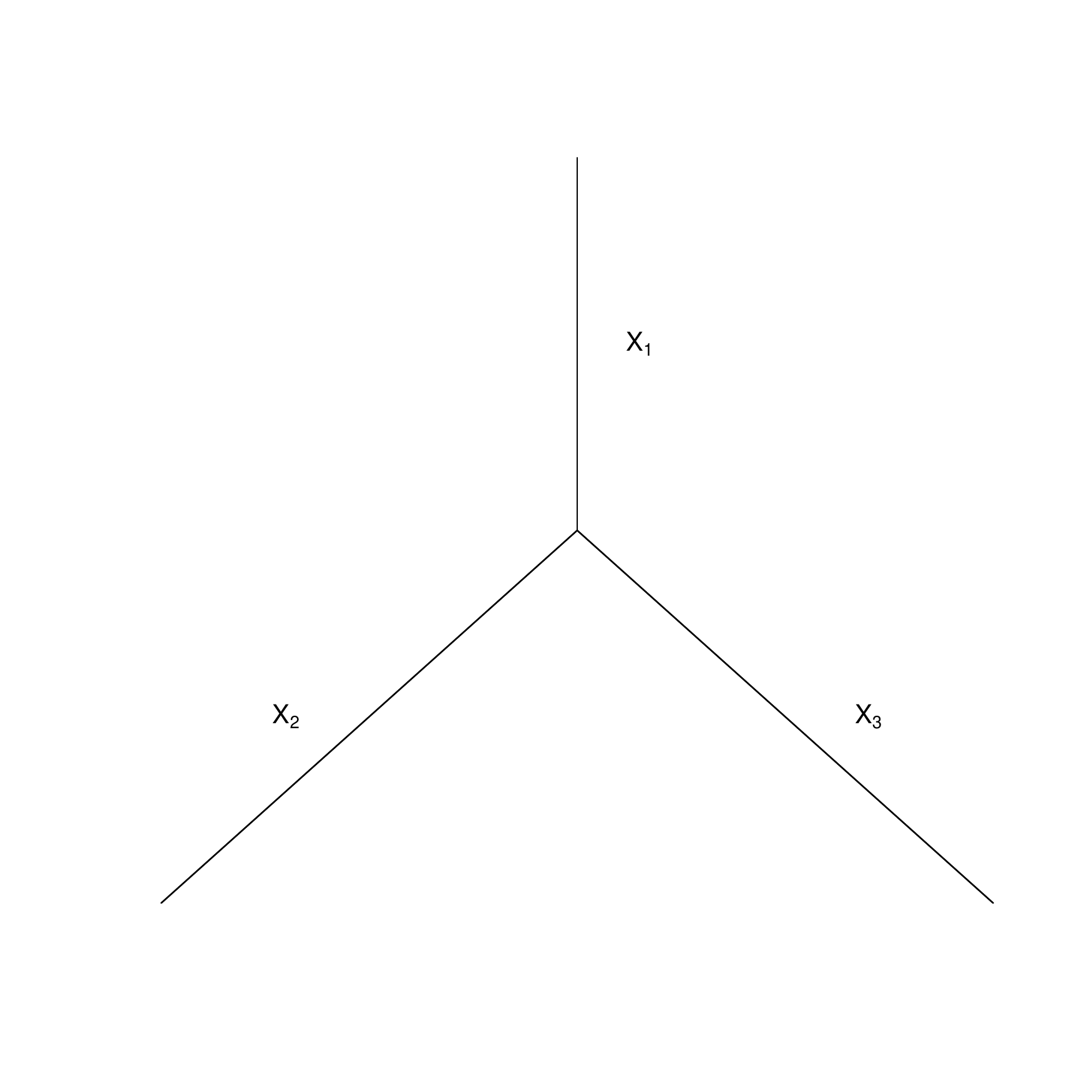}
    \caption{Simple network consisting of three links with transition times $X_i$, $i=1, 2, 3$.}
    \label{fig:network}
\end{figure}

Consider the simple network displayed in Figure~\ref{fig:network} with three links. Each link has a transition time $X_i$ distributed as $Exponential(\lambda_i)$ with mean $1/\lambda_i$.
The observed data from the network consist of the pairs $(Y_1, Y_2)$, where
$Y_1=X_1+X_2$ and
$Y_2=X_1+X_3$; $Y_1$ and $Y_2$ are clearly dependent.

We simulated $N=300$ observed pairs $(Y_1, Y_2)$ using $\lambda_1=0.3$, $\lambda_2=1/15$ and $\lambda_3=1/40$ and applied the SimILE method with
$nInt=50$ and $nSim=10^7$. We used the following priors:
$\lambda_1 \sim Lognormal(\log(0.3), 3)$,
$\lambda_2 \sim Lognormal(\log(1/15), 3)$,
$\lambda_2 \sim Lognormal(\log(1/40), 3)$.
To obtain posterior samples, we use the Metropolis-Hastings algorithm; we discard the first 1,000 samples as burnin and report on results from 10,000 subsequent samples. Plots of posterior draws of the three model parameters, not shown here, display good mixing.
The results for the SimILE method are given in Table~\ref{tab:networkPosterior}.

\begin{table}[H]
    \center
   \caption{Posterior Summaries for the Network Example}
    \label{tab:networkPosterior}
     \vspace{0.05 in}
    \renewcommand{\baselinestretch}{1}
\large\normalsize
    \begin{tabular}{llrrrrr}
        \hline
Method &Parameter&Mean&Std. Dev. & 2.5\% & 50\% & 97.5\%\\
 \hline
       SimILE& $\lambda_1$ &  0.404 & 0.112  & 0.254  & 0.387  & 0.650  \\ 
& $\lambda_2$ &  0.060  & 0.004  & 0.053  & 0.060  & 0.068  \\
& $\lambda_3$ &0.024  & 0.001  & 0.021  & 0.024  & 0.027 \\
        \hline
Exact& $\lambda_1$ &  0.397 & 0.102 & 0.255 & 0.378 & 0.654  \\
Likelihood& $\lambda_2$ &  0.060 & 0.004 & 0.052 & 0.060 & 0.068 \\
& $\lambda_3$ &  0.024 & 0.001 & 0.021 & 0.024 & 0.027 \\
        \hline

    \end{tabular}
    \renewcommand{\baselinestretch}{1.5}
\large\normalsize
\end{table}

In this example, the joint pdf of $(y_1, y_2)$ has a closed form expression \cite{Lawrence2007}. 
See the results using the exact likelihood in Table~\ref{tab:networkPosterior}. The results for the SimILE method are nearly the same.

\section{Briefly Exploring the SimILE Method}
\label{sec:study}

Here, we consider data distributed as $Lognormal(0,1)$ of size $N=50$. We use the priors: $\mu \sim Normal(0,10)$ and $\sigma \sim Lognormal(0,10)$.
Table~\ref{tab:studyPosterior} displays the posterior summaries for the exact likelihood based on 10,000 draws after 1,000 burnin draws. In the examples in Section~\ref{sec:examples}, we used $nInt=100$ and $nSim=10^7$ that performed well. This choice also works well here although $nSim=10^8$ performs somewhat better, whereas $nInt=50$ performs somewhat worse as does $nSim=10^6$. In applying the method, the practitioner can ballpark estimates for the model parameters (say, using $nInt=100$ and $nSIm=10^7$) and explore various $nInt$ and $nSim$ combinations like we do simulating  ``observed" data to identify $nInt$ and $nSim$ values to be used in the actual analysis.
For example,
$nInt$ can be chosen too large so that some of the observed data may have associated zero relative frequency intervals too often. Then, the MCMC algorithms will get stuck and not mix well.

\begin{table}[H]
    \center
   \caption{Posterior Summaries for the SimILE Method Exploration}
    \label{tab:studyPosterior}
     \vspace{0.05 in}
    \renewcommand{\baselinestretch}{1}
\large\normalsize
    \begin{tabular}{llrrrrr}
        \hline
Method &Parameter&Mean&Std. Dev. & 2.5\% & 50\% & 97.5\%\\
 \hline

Exact&$\mu$ & -0.253 &  0.150 & -0.544 & -0.249 &  0.033\\
Likelihood& $\sigma$ &  1.034 & 0.107 & 0.854 & 1.023 & 1.278 \\
        \hline  
        \hline
             \multicolumn{7}{c}{$nInt=100$, $nSim=10^8$} \\   
     \hline

SimILE& $\mu$ &  -0.254 &  0.147  & -0.546  & -0.253 &  0.029  \\
 & $\sigma$ & 1.029 & 0.108  & 0.839 & 1.023  & 1.257    \\
        \hline  
        \hline
             \multicolumn{7}{c}{$nInt=100$, $nSim=10^7$} \\   
     \hline

SimILE& $\mu$ &  -0.248 &  0.148 & -0.546 & -0.252 &  0.039 \\
 & $\sigma$ & 1.023 & 0.104 & 0.845 & 1.013 & 1.260   \\
       
        \hline
        \hline
             \multicolumn{7}{c}{$nInt=50$, $nSim=10^7$} \\   
     \hline

SimILE& $\mu$ &  -0.251 &  0.141  & -0.530  & -0.253  &  0.033 \\
 & $\sigma$ & 1.025  & 0.114  & 0.836  & 1.012  & 1.289   \\
       
        \hline
        \hline
             \multicolumn{7}{c}{$nInt=100$, $nSim=10^6$} \\   
     \hline

SimILE& $\mu$ & -0.253  & 0.144  & -0.536  & -0.253  &  0.040 \\
& $\sigma$ & 1.028  & 0.106  & 0.848  & 1.016  & 1.264   \\
       
        \hline

    \end{tabular}
    \renewcommand{\baselinestretch}{1.5}
\large\normalsize
\end{table}

\subsection{Reducing Computation Using a Gaussian Process Emulator}
\label{sec:GP}

Rather than estimating the discretized implicit likelihood each time to evaluate a candidate, i.e., 
the so-called brute force approach, 
we might 
use a Gaussian process emulator as is done with computer experiments
\cite{Santner2019}. 
That is, using a Latin Hypercube Sample (LHS) of the parameter space, we estimate the discretized implicit likelihood by simulation at the design specified by the LHS. Then assuming a Gaussian process (GP) model, we predict the discretized implicit likelihood using GP emulation, a kriging estimate based on maximum likelihood estimation. Here, we predict the cumulative discretized probabilities and use the principle component analysis approach for handling multiple responses as implemented in the R package mlegp \cite{Dancik2008}. 

We consider the same lognormal setup as discussed in Section~\ref{sec:study}.
We used a 500 point maximin LHS design on the square  $[ -2,2 ] ^2$ for $(\mu, \log(\sigma))$
and estimate the discretized implicit likelihood with $nInt=100$ by simulation using $nSim=10^7$ simulations.
We also used 20 principal components with the mlegp package.
As before we use 10,000 draws after 1,000 burnin draws.
Table~\ref{tab:posteriorLHS} displays the posterior summaries using GP emulation whose results are very similar to
those reported in Table~\ref{tab:studyPosterior}.

\begin{table}[H]
    \center
   \caption{Posterior Summaries for the SimILE Method Using Gaussian Process Emulation}
    \label{tab:posteriorLHS}
     \vspace{0.05 in}
    \renewcommand{\baselinestretch}{1}
\large\normalsize
    \begin{tabular}{lrrrrr}
        \hline
Parameter&Mean&Std. Dev. & 2.5\% & 50\% & 97.5\%\\
 \hline

$\mu$ & -0.250 &  0.146 & -0.537 & -0.250 &  0.031\\
$\sigma$ &  1.027 & 0.109 & 0.840 & 1.018 & 1.226 \\
        \hline

    \end{tabular}
    \renewcommand{\baselinestretch}{1.5}
\large\normalsize
\end{table}

\section{Discussion}
\label{sec:disc}

In this article, we have proposed discretizing continuous data into intervals and using simulated data to estimate the likelihood when the likelihood is implicit.
The use of relative frequencies to estimate interval probabilities is easy to explain. And the method can be used when the practitioner does not know how to evaluate the likelihood exactly. We have seen through the examples and a limited study that the SimILE method results are similar to the exact likelihood results; the SimILE method results are slightly more uncertain, i.e., slightly wider credible intervals. However, a practitioner may have a short deadline to analyze data and the SimILE method provides a viable and principled way to perform the analysis quickly when the likelihood is implicit. 
(Note that for the examples presented,  the exact likelihoods were buried in our Ph.D.\ theses and publications, and therefore would be unknown to most practitioners.) The practitioner can also use the SimILE method to handle complex multiple data source problems that involve a mixture of implicit and explicit likelihood components. Also,
while the SimILE method cannot handle a large dimensional response like ABC, it could be applied to the data summaries that ABC uses.

\section*{Acknowledgments}
We thank C.\ C.\ Essix for her encouragement and support. 

\subsection*{Author contributions}
All the authors shared equally in the work for this article.

\subsection*{Conflict of interest}
The authors declare no potential conflict of interests.

\bibliographystyle{dcu}
\bibliography{bib}

\end{document}